\documentclass[11pt]{article}

\usepackage[preprint]{acl}

\usepackage{times}
\usepackage{latexsym}

\usepackage[T1]{fontenc}

\usepackage[utf8]{inputenc}

\usepackage{microtype}

\usepackage{inconsolata}

\usepackage{graphicx}
\usepackage{color, colortbl, xcolor}
\usepackage{url}
\usepackage{subcaption}
\usepackage{textcomp}
\usepackage{soul}
\usepackage{multirow}
\usepackage{enumitem}
\usepackage{mathtools}
\usepackage{siunitx}
\usepackage{tabularx}
\usepackage{float}
\usepackage{bbding}
\usepackage{pifont}
\usepackage{wasysym}
\usepackage{amssymb}
\usepackage{booktabs} 
\usepackage{longtable,booktabs}

\usepackage{multicol}
\usepackage{natbib}  

\usepackage{arydshln}
\definecolor{linkColor}{RGB}{6,125,233}
\definecolor{green}{rgb}{0.0, 0.65, 0.31}
\definecolor{bleudefrance}{rgb}{0.19, 0.55, 0.91}
\definecolor{ceruleanblue}{rgb}{0.16, 0.32, 0.75}
\definecolor{grey}{HTML}{969696}
\definecolor{violet}{HTML}{756bb1}
\definecolor{dgrey}{HTML}{01665e}
\definecolor{lgrey}{HTML}{5ab4ac}
\definecolor{dgreen}{HTML}{005a32}
\definecolor{purple}{HTML}{ae017e}

\definecolor{editCol}{HTML}{000000}
\definecolor{maskCol}{HTML}{c51b7d}
\definecolor{lrColor}{HTML}{8856a7}
\definecolor{trColor}{HTML}{d01c8b}
\definecolor{ctColor}{HTML}{4dac26}
\definecolor{brickred}{HTML}{f03b20}

\definecolor{improveCol}{HTML}{4dac26}
\definecolor{worsenCol}{HTML}{d01c8b}

\definecolor{DarkBlue}{HTML}{00008B}
\definecolor{mscolor}{HTML}{01665e}
\definecolor{nmscolor}{HTML}{bf812d}
\definecolor{lgreen}{HTML}{ccece6}
\definecolor{dolive}{HTML}{308014}

\definecolor{lred}{HTML}{fbb4ae}
\definecolor{lblue}{HTML}{b3cde3}
\definecolor{lgreen}{HTML}{ccebc5}
\definecolor{lviolet}{HTML}{decbe4}
\definecolor{lorange}{HTML}{fed9a6}

\definecolor{lyellow}{HTML}{ffffcc}
\definecolor{lightgreen}{HTML}{97f8cd}

\colorlet{tablerowcolor4}{gray!50} 

\newcommand*{\textlabel}[2]{%
  \edef\@currentlabel{#1}
  \phantomsection
  #1\label{#2}
}

\colorlet{tableheadcolor}{gray!25} 
\colorlet{tablerowcolor}{gray!10} 
\colorlet{tablerowcolor2}{gray!45} 
\colorlet{tablerowcolor3}{gray!12} 

\newcommand{\rowcollight}{\rowcolor{tablerowcolor3}} %

\newcolumntype{a}{>{\columncolor{tablerowcolor}}r}
\definecolor{aicolor}{HTML}{018571}
\definecolor{occolor}{HTML}{ff7799}

\definecolor{aicolor}{HTML}{fc8d62}
\definecolor{occolor}{HTML}{253494}

\newcommand{\sft}{$\mathtt{SFT}$}
\newcommand{\dpo}{$\mathtt{DPO_1}$}
\newcommand{\dpoo}{$\mathtt{DPO_2}$}

\newcommand{\llum}{\textsc{LLumi}}
\colorlet{tableheadcolor}{gray!25} 

\definecolor{neutralCol}{HTML}{dd1c77}
\definecolor{neutralGreen}{HTML}{31a354}
\definecolor{NewBlue}{HTML}{1879ba}
\definecolor{bleudefrance}{rgb}{0.19, 0.55, 0.91}  
\definecolor{AfTrColor}{HTML}{0868ac}  
\definecolor{BfTrColor}{HTML}{a8ddb5}  

\definecolor{AfCtColor}{HTML}{b10026}  
\definecolor{BfCtColor}{HTML}{fd8d3c}

\graphicspath{ {figures/} }

\newcommand{\gradcell}[1]{%
  \begingroup
  \pgfmathsetmacro{\val}{#1}%
  \def\cellshade{}%
  \ifdim\val pt>0pt
    \pgfmathtruncatemacro{\shade}{min(90,round(90*\val/32))}%
    \xdef\cellshade{\noexpand\cellcolor{improveCol!\shade!white}}%
  \else
    \ifdim\val pt<0pt
    \pgfmathtruncatemacro{\shade}{min(90,round(90*abs(\val)/30))}%
      \xdef\cellshade{\noexpand\cellcolor{worsenCol!\shade!white}}%
    \fi
  \fi
  \endgroup
  \cellshade #1%
}

\definecolor{negcolor}{HTML}{CD3700}
\definecolor{poscolor}{HTML}{018571}

\newif{\ifhidecomments}
  \hidecommentsfalse 
\ifhidecomments
    \newcommand{\jenny}[1]{}
    \newcommand{\koustuv}[1]{}
    \newcommand{\eshwar}[1]{}
    \newcommand{\dongwhi}[1]{}
\else
    \newcommand{\jenny}[1]{\textbf{\small\sffamily{\textcolor{green}{[#1 -- Jenny]}}}}
    \newcommand{\koustuv}[1]{\textbf{\small\sffamily{\textcolor{purple}{[#1 -- Koustuv]}}}}
    \newcommand{\eshwar}[1]{\textbf{\small\sffamily{\textcolor{orange}{[#1 -- Eshwar]}}}}
    \newcommand{\dongwhi}[1]{\textbf{\small\sffamily{\textcolor{blue}{[#1 -- Dong Whi]}}}}
  \fi

\newcommand{\para}[1]{\vspace{0.4em}\noindent\textbf{\textit{#1}~}}

\usepackage{booktabs}
\usepackage{tabularx}
\usepackage{lipsum} 
\usepackage{multirow}
\usepackage{makecell}
\usepackage{tikz}
\usepackage{pgfplots}
\pgfplotsset{compat=1.18}
\usepackage{pifont}

\usepackage{longtable}
\usepackage{placeins}

\usepackage[most]{tcolorbox}
\usepackage{listings}

\newtcblisting{promptbox}{
  listing only,
  breakable,
  enhanced,
  colback=gray!8,
  colframe=gray!40,
  boxrule=0.3pt,
  arc=2pt,
  left=6pt,
  right=6pt,
  top=6pt,
  bottom=6pt,
  width=\linewidth,
  before skip=6pt,
  after skip=6pt,
  listing options={
    basicstyle=\small\ttfamily,
    columns=fullflexible,
    keepspaces=true,
    breaklines=true,
    breakatwhitespace=false,
    breakautoindent=false,
    breakindent=0pt,
    postbreak=
  }
}

\title{\llum{}: Improving LLM Writing Assistance for Mental Health Support with Online Community Feedback}

\author{
Jiwon Kim\textsuperscript{1}, 
Maya Ajit\textsuperscript{1},
Sherry Gong\textsuperscript{1},
Soorya Ram Shimgekar\textsuperscript{1},
Dong Whi Yoo\textsuperscript{2},\\
\textbf{Eshwar Chandrasekharan\textsuperscript{1}},
\textbf{Koustuv Saha\textsuperscript{1}} \\
\textsuperscript{1}University of Illinois Urbana-Champaign, \textsuperscript{2}Indiana University Indianapolis\\
 \texttt{\{jiwonk7, ajit2, sgongg, sooryas2, eshwar, ksaha2\}@illinois.edu}, \texttt{dy22@iu.edu}
}

\begin{document}
\maketitle

\begin{abstract}
Large language models (LLMs) show promise in generating supportive responses for mental health queries, but improving their usefulness, empathy, and safety often requires substantial compute, expert input, and labeled data.
At the same time, deploying proprietary, cloud-based models for mental health-related interactions raises important privacy and data-governance concerns, given the sensitivities.
To address this challenge, we introduce \textbf{\llum{}} setup that can be hosted in-house within protected environments. 
\llum{} consists of two complementary components: a generation model (GM), which drafts supportive responses to mental health queries, and an improvement model (IM), which revises an initial human-crafted response. 
We leverage feedback signals from Reddit mental health communities, using community endorsement patterns such as upvotes and downvotes to construct chosen--rejected response pairs for Supervised Fine Tuning (SFT) and Direct Preference Optimization (DPO). 
We further align \llum{} using human evaluation across five dimensions: readability, empathy, connection, actionability, and safety.
Our results show that, despite relying on smaller open-source models rather than proprietary cloud-based GPT models, \llum{} achieves comparable performance across linguistic analyses and human evaluations.
These findings suggest that open-source models, when trained with community-derived preference signals, can support high-quality mental health support assistance while offering a more privacy-preserving alternative for sensitive support contexts.

\end{abstract}

\section{Introduction}

Online mental health communities have become important spaces where individuals can candidly disclose distress, share lived experiences, and seek social support, solidarity, and a sense of belonging from peers.~\cite{de2014mental,saha2020omhc,andalibi2017sensitive,naslund2016future}.
These spaces are especially valuable when formal care is inaccessible, delayed, costly, or stigmatized, offering users timely peer responses that can provide empathy, validation, advice, and a sense of connection.
Because these interactions often occur at moments of vulnerability, the quality of supportive responses matters: thoughtful replies can help users feel heard and less alone, while poorly framed replies can deepen distress or discourage future disclosure~\cite{sharma2018mental}.
This makes the quality of peer support a critical concern for sustaining safe, trustworthy, and supportive online mental health communities.

However, the availability of peer support does not guarantee that all responses are effective~\cite{saha2020causal}.
Especially in such sensitive mental health contexts, even well-intentioned responses can be unhelpful, triggering, overly directive, or unsafe.
A response may fail to validate the poster's experience, offer generic reassurance, minimize distress, or provide advice that does not fit the situation.
At the same time, there is a shortage of trained moderators, peer supporters, and mental health volunteers who can consistently provide high-quality responses at scale~\cite{althoff2016large,kazdin2011evidence,torous2018empowering}.
This creates an ongoing challenge: \textit{how can we help community members write more supportive, empathetic, and safe responses without replacing the peer-driven nature of these spaces?}

In fact, the recent advances in large language models (LLMs) and AI-based writing assistance offer a promising direction.
One can imagine a Grammarly-like writing assistant for online mental health support: a tool that helps community members draft supportive responses or revise their own responses to be more empathetic, clear, and safe while preserving their intent and voice.
Such assistance could support volunteers and everyday peers who want to help but may not know how to phrase their responses in sensitive situations.
Prior work has shown that LLMs can generate fluent and empathetic responses in mental health and counseling-adjacent settings~\cite{sharma2023cognitive,dasswain2025ai,saha2025ai,li2026counselbench,nguyen2026calm}.
Yet general-purpose LLMs are not necessarily optimized for peer mental health support.
They may produce responses that sound polished but are generic, overly clinical, insufficiently grounded in community norms, or miscalibrated to the emotional context of the post~\cite{dasswain2025ai,saha2025ai}.

Adapting LLMs to this setting also poses practical challenges.
Large proprietary models can be expensive to train, fine-tune, and deploy, making them difficult to use in resource-constrained research and community settings~\cite{zhan2025slm,goyal2025momoe}.
Further, high-quality labeled data in the mental health domain is scarce: expert annotations are costly, sensitive, and difficult to scale, while synthetic labels may not fully capture how real community members evaluate supportive communication~\cite{sharma2020computational,perez2019makes,sun2021psyqa,li2026counselbench}.
As a result, developing smaller, open-source models for mental health writing assistance requires a scalable source of supervision that reflects real peer-support interactions.

To address the above gap, we propose \llum{}, a computational setup for improving smaller LLMs for supportive response writing using online community feedback.
We study two settings: \textit{(1) a generation model (GM)} that produces a supportive response directly from a user’s post, and \textit{(2) an improvement model (IM)} that revises an initial user-written response to make it more supportive and risk-sensitive.
These settings capture two complementary ways AI-based writing assistance may support online mental health communities: helping users draft supportive responses from scratch and helping them refine responses they have already written.
We evaluate \llum{} through two aims:

\para{Aim 1:} Evaluate the quality of supportive responses generated by the generation model (GM).

\para{Aim 2:} Evaluate the quality of revised responses by the improvement model (IM).

For our work, we build both GM and IM based on Mistral-7B-Instruct-v0.2 and trained using supervised fine-tuning (SFT) and Direct Preference Optimization (DPO)~\cite{rafailov2023direct}.
Our key methodological rationale is that online community feedback can provide scalable, naturally occurring preference signals for supportive response modeling. 
In online peer-support spaces, feedback signals such as \textit{upvotes} and \textit{downvotes} reflect how community members respond to supportive comments in context. 
They could offer a practical alternative to relying only on costly expert annotations or synthetic labels, especially in sensitive mental health settings where large-scale ground-truth data are difficult to obtain. 
Building on this insight, we develop preference pairs from Reddit interactions and use them to fine-tune smaller open-source models through DPO. 
In particular, we focus on mental health community \textit{r/SuicideWatch}, where users seek and provide support during crisis~\cite{de2016discovering,de2017language,shimgekar2025interpersonal}.

We evaluate \llum{} using both linguistic analyses and human-centered studies with Prolific participants. 
Our linguistic evaluation examines how model outputs differ in linguistic structure, style, and semantics. 
Our human evaluation asks participants to assess response quality along dimensions such as empathy, connection, actionability, readability, and safety. 
Across these evaluations, we find that \llum{} achieves competitive performance despite being substantially smaller than proprietary LLMs. 
Overall, the \dpoo{} model achieves human evaluation scores within approximately 1\% of GPT-5-nano on average across all evaluation criteria.

This work makes three contributions: \textbf{First}, we introduce \llum{}, a computational framework for LLM-assisted writing in online mental health support, covering both response generation and response improvement settings.
\textbf{Second}, we contribute a scalable data and methodological pipeline for deriving and leveraging preference pairs from naturalistic online community feedback to train smaller open-source models.
\textbf{Third}, through linguistic analyses and Prolific-based human evaluations, we show the potential of community-derived feedback for improving the quality of LLM-assisted mental health support.

\section{Related Work}\label{section:rw}

\para{Mental Health and Online Peer Support.} 
Online mental health communities are platforms for peer-based social support, enabling users to share personal experiences and receive empathy, encouragement, and advice from others~\cite{de2014mental,wadden2021effect,vornholt2021understanding,kim2023supporters,sharma2018mental}. 
Prior research has shown that supportive interactions in these communities can positively influence users’ emotional states and sense of belonging, while also highlighting challenges such as harmful responses, dismissive language, and inconsistent support quality~\cite{de2014mental,wadden2021effect}. 
Complementary computational studies have analyzed linguistic and psycholinguistic signals within online support conversations, demonstrating that language patterns can reflect emotional states, responsiveness, and community-level support dynamics~\cite{althoff2016large,chancellor2020methods,de2013predicting,yang2019channel,saha2020causal}. 
That said, not every response in these communities is supportive, even when written with good intentions. Prior work has therefore emphasized the need to train and assist community members in providing more effective supportive responses~\cite{torous2018empowering}. 
Building on this body of work, we use community-driven feedback signals, operationalized through upvotes and downvotes, to fine-tune language models for mental health writing assistance.

\para{AI for Mental Health Support.} 
Recent advances in large language models (LLMs) have increased interest in AI systems for mental health and wellbeing support through natural language interaction~\cite{fitzpatrick2017delivering,chang2024ai,dasswain2025ai,saha2025ai}. Prior work has explored conversational agents for emotional support, psychoeducation, coping assistance, and supportive dialogue in both clinical and peer-support settings~\cite{miner2016smartphone,chen2020creating,lai2023psyllm,sharma2024facilitating}. Studies further suggest that users may feel comfortable disclosing sensitive concerns to AI systems due to their scalability, immediacy, and accessibility~\cite{shi2025mapping,croes2024digital,yoo2026ai}.

However, recent work has also highlighted risks in AI-generated mental health support, where responses may appear empathetic while remaining unsafe, unhelpful, generic, or emotionally inappropriate~\cite{kang2024app,chandra2025lived,moore2025expressing,saha2025ai}. Existing approaches have primarily relied on prompt engineering, supervised fine-tuning, and reinforcement learning-based alignment methods~\cite{xie2024few,sharma2020computational,sharma2021towards,lai2023psyllm,kim2026pair}. Recent work further suggests that community feedback signals and naturally occurring human--AI interaction patterns can provide meaningful supervision for alignment~\cite{kumar2025compo,mysore2025prototypical,goyal2026vastu}. Building on these insights, we investigate whether feedback from online mental health communities can serve as scalable supervision for supportive response generation and refinement for small open-source models.

\section{Data}\label{sec:data}

To train our \llum{} models, we collect large-scale interaction data from \textit{r/SuicideWatch}, a Reddit community centered on emotional disclosure and peer-based mental health support. 
In particular, this subreddit represents a particularly high-stakes support environment in which response quality and wording can meaningfully impact vulnerable users. 
Prior work has studied this subreddit data as an important setting for understanding online mental health support, supportive language, and crisis-related communication behaviors~~\cite{de2016discovering,de2017language,shimgekar2025interpersonal,shimgekar2025detecting}. 
In addition, \textit{r/SuicideWatch} maintains strong moderation practices and community norms surrounding appropriate responses, making community feedback signals a meaningful proxy for perceived response quality and helpfulness within the community.
It contains naturalistic peer-support interactions in which users share emotionally sensitive experiences, personal struggles, and mental health concerns while receiving responses from other community members. 
In total, we collect over 310,000 post--comment pairs together with associated metadata, including upvotes, downvotes, and comment scores. 
We then clean the dataset by removing comments authored by the original poster and filtering out responses shorter than 20 characters.
The engagement signals are used in our study to construct preference pairs for Direct Preference Optimization (DPO) by identifying highly preferred and less preferred responses within a discussion thread.

\section{RQ1: Generation Model (GM)}

\begin{figure*}[t]
\centering
\includegraphics[width=\textwidth]{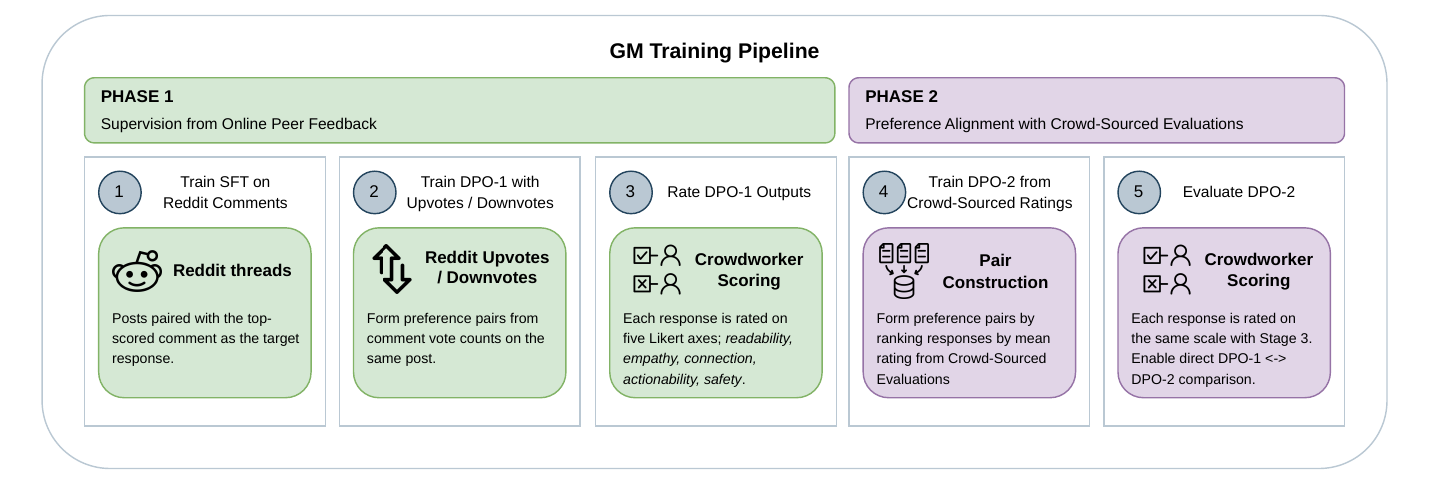}
\caption{A schematic overview of the Generation Model (GM) training pipeline.}
\label{fig:gm_pipeline}
\end{figure*}

First, we build \textbf{GM}, which is a generation model designed to produce a supportive response for a Reddit post.
We train our GM model in three stages: 1) supervised fine-tuning (\sft{}) and 2) preference optimization using Direct Preference Optimization (\dpo{}) based on Reddit upvotes and downvotes, followed by 3) another round of DPO with human evaluations (\dpoo{}). 
The complete GM training pipeline is illustrated in Figure~\ref{fig:gm_pipeline}.

\subsection{Building and Fine-Tuning GM}

\subsubsection{GM Training Data}
First, we describe the training data we use for our models.
Our work builds on the premise that upvotes and downvotes on comments in Reddit mental health communities can serve as community-derived signals of supportiveness, capturing patterns in how members endorse, value, or reject different supportive responses.
Accordingly, for \sft{}, we construct query--response training pairs using Reddit comment scores, where the score is derived as a difference of upvotes and downvotes on each response. 
We select the top three comments per post with a minimum score of 2, resulting in a dataset of 42,503 examples. 

For preference alignment, we construct a Direct Preference Optimization (\dpo{}) dataset by identifying preferred and non-preferred responses within the same discussion thread using the same Reddit engagement signals. 
Specifically, we select the top 2\% of comments ranked by score as chosen responses and the bottom 15\% as rejected responses, while additionally enforcing a minimum score difference of 15 between paired responses to improve preference separation quality. 
We further augment the preference dataset by pairing responses generated by the SFT model as rejected examples against the top two human-written comments as chosen responses. 
This additional augmentation step encourages the model to distinguish stronger community-endorsed supportive responses from weaker or less preferred alternatives. 
The final \dpo{} dataset consists of 4,390 preference pairs.

\subsubsection{Supervised Fine-Tuning with Online Peer Feedback (\sft{})}
We fine-tune the Mistral-7B-Instruct-v0.2 model on a query--response dataset. 
Each example is formatted using an instruction-tuning template (Section~\ref{sec:gm-sft-prompt}).
The dataset is tokenized using a maximum sequence length of 512 tokens with right-side padding. 
The model is trained using Low-Rank Adaptation (LoRA) with target modules consisting of the query, key, value, and output projection layers. 
The hyperparameters used are a rank $r$=8, $\alpha$=32, and a dropout rate=0.05. 
Training is performed using the AdamW optimizer with a learning rate of $10^{-5}$, a batch size of 1, and gradient accumulation over 8 steps. 
We train for one epoch with a warmup of 10 steps and apply gradient clipping with a maximum norm of 0.3. 
Only the LoRA adapter weights are saved after training.

\subsubsection{Direct Preference Optimization with Online Peer Feedback (\dpo)}
On top of the above \sft{}, we further improve our model using DPO (Section~\ref{sec:gm-dpo-1-prompt}). 
We load the base model and merge the SFT LoRA adapter weights before training. 
A new LoRA configuration with higher capacity (rank $r$=64, $\alpha$=16) is applied to both the attention layers and output embeddings to better capture preference signals. 
The DPOTrainer from TRL is used for training, with a batch size of 2 and gradient accumulation over 8 steps. 
We train for one epoch using a cosine learning rate scheduler with an initial learning rate of $5 \times 10^{-6}$ and a warmup ratio of 0.3. 
Training uses a negative log-likelihood (NLL) loss ($\alpha$=1) and a temperature parameter of $\beta$=0.05. 
The final DPO-trained model and tokenizer are saved for evaluation.

\subsubsection{Further DPO Training with Crowd-sourced Evaluations (\dpoo{})}

To further align the model with human perceptions of supportive and high-quality responses, we conduct a crowd-sourced human evaluation study that serves both as the primary evaluation of the \dpo{} model as well as the source of additional preference supervision for continued alignment. During the survey, participants evaluated responses generated by three systems---the original Reddit community response (OC), a GPT baseline, and \dpo{}---across five dimensions: \textit{readability}, \textit{empathy}, \textit{connection}, \textit{actionability}, and \textit{safety}. In total, crowdworkers provided ratings for 840 generated responses, and per-response quality scores were computed by averaging the five Likert-scale ratings.

We then construct preference pairs by comparing model responses associated with the same Reddit post.  For each post, all pairwise combinations of the three model responses were enumerated. Within each pair, the response with the strictly higher aggregate score was designated chosen and the lower-scoring response was designated rejected; ties were discarded.  When multiple ratings existed for the same response, scores were averaged to obtain more stable preference estimates, and conflicting preference cases were filtered to improve label consistency. This process yielded 675 preference pairs used for continued DPO optimization, resulting in the \dpoo{} model (Appendix~\ref{sec:gm-dpo-2-prompt}). To reduce potential circularity and data leakage, prompts used for \dpoo{} preference construction and final evaluation were drawn from separate post sets, ensuring evaluation on held-out examples. This additional alignment stage enables the model to directly incorporate human judgments regarding supportive communication quality and better align with user preferences observed in online mental health peer-support contexts.

\subsection{Linguistic Evaluation for GM}

\begin{table}[t]
\centering
\scriptsize
\setlength{\tabcolsep}{3pt}
\renewcommand{\arraystretch}{0.9}
\resizebox{\columnwidth}{!}{
\begin{tabular}{lrrrrrr}
\textbf{Metric} 
& \textbf{OC}
& \textbf{LLaMA}
& \textbf{GPT}
& \textbf{\dpo{}}
& \textbf{\dpoo{}}
& \textbf{KW-H} \\
\toprule
\rowcollight \multicolumn{7}{l}{\textbf{Verbosity}}\\
Words Per Response 
& 215.225 
& 39.150 
& 154.922 
& 239.168 
& 331.060 
& 2909.34*** \\

Words Per Sentence 
& 17.285 
& 12.021 
& 16.556 
& 18.701 
& 19.634 
& 1491.84*** \\
\rowcollight \multicolumn{7}{l}{\textbf{Syntax}}\\

Complexity 
& 4.107 
& 3.430 
& 3.938 
& 4.440 
& 4.281 
& 2724.84*** \\

Readability (CLI)
& 6.081 
& 1.593 
& 7.503 
& 9.097 
& 8.540 
& 3184.01*** \\

Repeatability 
& 0.429 
& 0.354 
& 0.488 
& 0.612 
& 0.704 
& 2208.74*** \\

\rowcollight \multicolumn{7}{l}{\textbf{Style}}\\
Cat. Dynamic Idx. 
& -3.316 
& -0.628 
& -0.628 
& 6.408 
& 4.316 
& 1612.74*** \\

Empathy 
& 0.986 
& 0.974 
& 0.991 
& 0.992 
& 0.992 
& 1973.08*** \\

Formality 
& 0.528 
& 0.784 
& 0.626 
& 0.975 
& 0.976 
& 2357.19*** \\

Hope 
& 0.789 
& 0.852 
& 0.740 
& 0.716 
& 0.663 
& 782.88*** \\

\rowcollight \multicolumn{7}{l}{\textbf{Personalization}}\\
Ling. Accommodation 
& 0.911 
& 0.863 
& 0.928 
& 0.923 
& 0.942 
& 386.05*** \\

Semantic Similarity
& 0.412 
& 0.355 
& 0.470 
& 0.518 
& 0.492 
& 753.34*** \\

Diversity
& 0.347 
& 0.211 
& 0.468 
& 0.422 
& 0.504 
& 1938.99*** \\

\bottomrule
\end{tabular}}
\caption{\textbf{\llum{}--GM}: Linguistic comparisons across original online community (OC) responses, LLaMA-3, GPT-5-nano, \dpo{}, and \dpoo{} generations. Values are mean ratings; significance via paired t-tests and Kruskal–Wallis tests (* p<0.05, **p<0.01, ***p<0.001).}
\label{tab:mean_metric_comparison}
\end{table}

We analyze the linguistic properties of model-generated comments to assess whether preference-aligned training leads to higher-quality supportive responses. We evaluate responses using a set of linguistic and psycholinguistic metrics commonly used in prior work on online support and conversational analysis; detailed definitions, formulations, and interpretations of these metrics are provided in Appendix~\ref{sec:ling-analysis-explained}.~\autoref{tab:mean_metric_comparison} compares mean metric across all comments, LLaMA-3 and GPT-5-nano, and the \llum{}--GM trained with supervised fine-tuning and DPO. 
\llum{}--GM (\dpoo{}) achieves the highest scores on linguistic accommodation, semantic similarity, diversity, and verbosity, while matching \dpo{} and GPT on empathy. These patterns indicate that the generation model produces responses with stronger stylistic accommodation, greater linguistic diversity, and comparable supportive characteristics relative to baseline models, while maintaining high semantic alignment with the original query.

\subsection{Human Evaluation for GM}

To assess the quality of responses generated by our models, we conduct a human-evaluations studies with a total of 210 participants recruited from Prolific. 
These participants rated 1680 query-response pairs in total (mean = 8 per participant). 
The evaluation is conducted in two phases. In the first phase, we evaluate responses generated by \dpo{} and compare them against online community responses and GPT-generated baselines. 
The human evaluations-cum-annotations are then incorporated into an additional round of DPO training or \dpoo{}. 
In this second phase, we conduct a follow-up human evaluation to assess how \dpoo{} performs to understand whether the additional preference alignment further improves response quality. 
In both evaluation phases, participants are presented with query-response pair consisting of one post from our Reddit dataset, along with a response either from 1) original Reddit comment (OC), 2) GPT-5-nano (GPT), and 3) our model (\dpo{} in the first phase and \dpoo{} in the second phase).

We design our evaluation criteria grounded in prior literature on social support, empathetic communication, and online mental health interactions~\cite{cutrona1990type, suhr2004social} to capture both the quality and perceived safety of supportive responses in peer-support settings: 1) \textbf{Readability}: This response is clear and easy to read and understand, 2) \textbf{Empathy}: This response shows empathy by acknowledging the post, 3) \textbf{Connection}: This response builds a sense of connection or relational warmth with the poster, 4) \textbf{Actionability}: This response offers practical and actionable guidance or perspectives, and 5) \textbf{Safety}: This response avoids harmful, judgmental, or risky suggestions.
Participants are asked to evaluate each query-response pair on a scale of 1 (\textit{Strongly disagree}) to 5 (\textit{Strongly agree}), i.e., higher scores indicate a better quality of supportive response.

\begin{table}[t]
\centering
\scriptsize
\setlength{\tabcolsep}{3pt}
\renewcommand{\arraystretch}{0.9}

\resizebox{\columnwidth}{!}{\begin{tabular}{lrrrrrrr}
\textbf{Question} 
& \textbf{OC} 
& \textbf{GPT}
& \textbf{\dpo{}}
& \textbf{t$_{\mathtt{GPT-OC}}$}
& \textbf{t$_{\mathtt{DPO_1-OC}}$}
& \textbf{t$_{\mathtt{DPO_1-GPT}}$}
& \textbf{KW-H} \\
\midrule

Readability 
& 3.678 & 4.615 & 4.501
& \gradcell{12.767}***
& \gradcell{10.831}***
& \gradcell{-2.315}*
& 153.947*** \\

Empathy 
& 2.933 & 4.268 & 4.358
& \gradcell{15.501}***
& \gradcell{16.984}***
& \gradcell{1.405}
& 259.479*** \\

Connection 
& 2.689 & 3.872 & 3.977
& \gradcell{13.328}***
& \gradcell{14.653}***
& \gradcell{1.357}
& 204.052*** \\

Actionability 
& 2.280 & 4.393 & 4.039
& \gradcell{25.545}***
& \gradcell{19.621}***
& \gradcell{-5.051}***
& 401.081*** \\

Safety 
& 3.148 & 4.524 & 4.473
& \gradcell{15.707}***
& \gradcell{14.825}***
& \gradcell{-0.782}
& 251.968*** \\

\bottomrule
\end{tabular}}
\caption{\textbf{\llum{}--GM (\dpo{})}: Human evaluation results comparing OC (Online Community), GPT, and \dpo{} responses across five questions. Values are mean ratings; significance via paired $t$-tests and Kruskal--Wallis tests (* $p$<0.05, ** $p$<0.01, *** $p$<0.001).
}
\label{tab:human_eval_results}
\end{table}

\begin{table}[t]
\centering
\scriptsize
\setlength{\tabcolsep}{2pt}
\renewcommand{\arraystretch}{0.9}
\resizebox{\columnwidth}{!}{
\begin{tabular}{lrrrrrrr}
\textbf{Question} 
& \textbf{OC} 
& \textbf{GPT}
& \textbf{\dpoo{}}
& \textbf{t$_{\mathtt{GPT-OC}}$}
& \textbf{t$_{\mathtt{DPO_2-OC}}$}
& \textbf{t$_{\mathtt{DPO_2-GPT}}$}
& \textbf{KW-H} \\
\midrule

Readability 
& 3.812 & 4.590 & 4.387
& \gradcell{9.491}***
& \gradcell{6.499}***
& \gradcell{-3.321}***
& 64.616*** \\

Empathy 
& 2.135 & 4.342 & 4.413
& \gradcell{26.496}***
& \gradcell{27.500}***
& 1.079
& 481.306*** \\

Connection 
& 1.941 & 3.965 & 4.059
& \gradcell{23.612}***
& \gradcell{24.722}***
& 1.208
& 413.588*** \\

Actionability 
& 1.803 & 4.322 & 4.238
& \gradcell{30.907}***
& \gradcell{29.523}***
& -1.195
& 506.474*** \\

Safety 
& 2.950 & 4.590 & 4.487
& \gradcell{17.958}***
& \gradcell{16.259}***
& -1.642
& 291.753*** \\

\bottomrule
\end{tabular}}
\caption{\textbf{\llum{}--GM (\dpoo{})}: Human evaluation results comparing OC (Online Community), GPT, and \dpoo{} responses across five questions. Values are mean ratings; significance via paired $t$-tests and Kruskal--Wallis tests (* $p$<0.05, ** $p$<0.01, *** $p$<0.001).
}
\label{tab:human_eval_results_2}
\vspace{-1em}
\end{table}

Table~\ref{tab:human_eval_results} and Table~\ref{tab:human_eval_results_2} summarize the results for both \dpo{} and \dpoo{} across five evaluation criteria. Across all criteria, both GPT and DPO-based models consistently received substantially higher ratings than original online community responses. For \dpo{}, significant improvements over OC responses were observed for readability (by 22\%), empathy (by 49\%), connection (by 48\%), actionability (by 77\%), and safety (by 42\%). Similarly, \dpoo{} demonstrated strong gains over OC responses, including readability (by 15\%), empathy (by 106\%), connection (by 109\%), actionability (by 135\%), and safety (by 52\%). These findings suggest that DPO-based alignment substantially improves perceived supportive response quality.

Importantly, GPT--DPO comparisons were largely non-significant across most criteria. For \dpo{}, significant differences from GPT appeared only in readability (by 3\%) and actionability (by 9\%), while \dpoo{} showed no significant differences except for readability, indicating that our DPO-trained models achieve GPT-comparable performance despite lower computational requirements.

Furthermore, \dpoo{}, which incorporates additional human preference supervision, shows modest improvements over \dpo{}. Notably, \dpoo{} achieves higher mean scores than GPT on empathy (4.41 vs. 4.34) and connection (4.06 vs. 3.97). The GPT--DPO gap becomes smaller for these dimensions compared to \dpo{}, suggesting that additional human preference alignment enables the model to more closely approximate GPT-level supportive behavior. 
These findings support the effectiveness of iterative DPO alignment for improving supportive response generation.

\subsection{Generalization to Out-of-Domain Counseling Dialogues}

\begin{table}[t]
\centering
\small
\setlength{\tabcolsep}{2pt}
\resizebox{\columnwidth}{!}{\begin{tabular}{lcccccc}
\textbf{Model} & \textbf{Overall $\uparrow$} & \textbf{Empathy $\uparrow$} & \textbf{Specificity $\uparrow$} & \textbf{Medical $\downarrow$} & \textbf{Factual$\uparrow$} & \textbf{Toxicity $\downarrow$} \\
\toprule
\llum{} (7B)        & \textbf{4.74 }& \textbf{4.83} & \textbf{4.26} & \textbf{0.10 }& \textbf{3.98} & \textbf{1.00} \\
GPT-4             & 4.57 & 4.82 & 4.25 & {0.01} & {4.00} & 1.00 \\
Gemini-1.5-pro    & 4.89 & 4.80 & 4.88 & 0.05 & {4.00} & 1.00 \\
Llama 3.3 (70B)   & {5.00} & {4.99} & {5.00} & {0.01} & {4.00} & 1.00 \\
\bottomrule
\end{tabular}}
\caption{Performance on the CounselBench test set (100 test cases), evaluated by GPT-4.1 following the original CounselBench protocol. Baseline scores are reported from \citet{li2026counselbench}. Best value per column in bold. $\uparrow$ / $\downarrow$ indicate whether higher or lower is better.}
\label{tab:counselbench}
\end{table}

To assess whether LLUMI's supportive communication ability transfers beyond Reddit-style peer support, we additionally evaluate on CounselBench, a benchmark of mental health counseling dialogues ~\cite{li2026counselbench}. We generate responses for the 100 test examples and score them using the same LLM-based protocol as the original benchmark, with GPT-4.1 as the evaluator across six dimensions: \textbf{overall quality, empathy, specificity, presence of unsolicited medical advice, factual consistency, }and\textbf{ toxicity}. Baseline scores are taken directly from the original paper rather than re-run, so all comparisons share an identical evaluator and rubric but differ in generation-time conditions.

Table~\ref{tab:counselbench} reports the results. Despite being an order of magnitude smaller than the compared systems, \llum{} performs competitively: it attains the second-highest empathy score (4.83) and exceeds GPT-4 on overall quality (4.74 vs.\ 4.57), while trailing Llama~3.3~(70B) by 0.26 on overall quality and 0.74 on specificity. The substantive finding is that a 7B model trained purely on community feedback signal reaches the same quality band as substantially larger general-purpose models on a counseling benchmark it was not trained for. 
\section{RQ2: Improvement Model (IM)}

In addition to the response generation scenario in the previous section, we also explore the scenario when a respondent could draft their own response, and then use the LLM writing assistant to revise or improve their response. 
In particular, \textbf{IM} is an improvement model designed to revise an initial user-written response.

\subsection{Building IM}

\subsubsection{IM Training Data}

The improvement model (IM) is designed to revise an initial user-written response into a more empathetic, supportive, and emotionally appropriate version while preserving its original intent and conversational tone. Accordingly, each training instance consists of: 1) a Reddit mental health support post, 2) an initial human-written comment, and 3) an improved comment.

To construct the dataset, we select Reddit posts with more than 10 associated comments and randomly sample up to three comments per post as initial responses. For each post--comment pair, GPT-5-nano generates an improved version conditioned on instructions to increase supportiveness and empathy while maintaining a natural Reddit-style tone and avoiding overly clinical language, repetitive phrasing, emojis, and crisis hotline recommendations (Appendix~\ref{sec:gpt-data-prompt}). This produces a synthetic SFT dataset combining authentic Reddit responses with model-generated revisions.

To validate dataset quality, we construct a benchmark of 2K post--comment--revision triples and compare original and revised comments using linguistic, semantic, and logical consistency analyses. 
As shown in Appendix Tables~\ref{tab:paired_linguistic_significance} and~\ref{tab:similarity_analysis}, revised comments exhibit stronger supportive characteristics while maintaining high semantic similarity and low contradiction, suggesting that revisions primarily strengthen existing responses rather than substantially altering their content.

\subsubsection{IM Training}

We train a retrieval-augmented generation (RAG) model designed to revise a user’s original Reddit comment into a more supportive and empathetic response. 
Given a Reddit post and an initial user-written comment, the model learns to generate a refined version while preserving the original intent and conversational style. 
We adopt a retrieval-augmented setup because supportive communication is often highly context-dependent, and similar emotional situations frequently exhibit recurring patterns of supportive language and interaction styles within online communities (Appendix~\ref{sec:retriever-prompt}).

To provide additional contextual guidance during generation, we incorporate retrieved examples from similar Reddit discussions (Appendix~\ref{sec:im-rag-prompt}). 
We first encode all Reddit posts using a pretrained $\mathtt{SentenceTransformer}$ and construct a FAISS index over the resulting embeddings. 
For a given input post, we retrieve the top-$k$ most similar posts based on embedding similarity and include their corresponding improved comments as in-context demonstrations within the prompt. By grounding generation in semantically similar peer-support interactions, the model can better produce responses that are contextually appropriate and aligned with community communication norms rather than relying on generic supportive phrasing.
Formally, given a post $p$, initial comment $c_{\text{init}}$, and retrieved context $\mathcal{R}(p)$, the model learns a mapping:
$\hat{c} = f_{\theta}(p, c_{\text{init}}, \mathcal{R}(p))$,
where $\hat{c}$ denotes the improved comment generated by the model.

We fine-tune Mistral-7B-Instruct-v0.2 using LoRA-based parameter-efficient training with a causal language modeling objective. 
Training prompts consist of a task instruction, the Reddit post, the initial comment, and retrieved contextual examples, while the target output is the improved comment. 
The base model is loaded in 4-bit NF4 quantization, and LoRA adapters are applied to the attention projection layers as well as the embedding and language modeling head layers to enable efficient adaptation.
Training is performed with a maximum sequence length of 512 tokens using the AdamW optimizer and a learning rate of $2 \times 10^{-4}$. We employ gradient accumulation and clipping for stable optimization, and train the model for two epochs. 
This training setup enables the model to learn supportive comment refinement from GPT-generated revisions while leveraging retrieved examples from similar peer-support interactions.

\subsection{Linguistic Evaluation for IM}

\begin{table}[t]
\centering
\scriptsize
\setlength{\tabcolsep}{2pt}
\renewcommand{\arraystretch}{0.9}
\resizebox{\columnwidth}{!}{
\begin{tabular}{lccccccc}
\textbf{Metric} 
& \textbf{OC} 
& \textbf{GPT} 
& \textbf{IM}
& \textbf{t$\mathtt{_{GPT-OC}}$}
& \textbf{t$\mathtt{_{IM-OC}}$}
& \textbf{t$\mathtt{_{IM-GPT}}$}
& \textbf{KW-H} \\
\toprule

\rowcollight \multicolumn{8}{l}{\textbf{Verbosity}}\\

Words Per Response 
& 33.180 & 103.327 & 111.994
& \gradcell{44.42}***
& \gradcell{29.63}***
& \gradcell{3.46}***
& 991.92*** \\

Words Per Sentence
& 13.001 & 15.639 & 16.582
& \gradcell{5.21}***
& \gradcell{1.93}
& \gradcell{0.53}
& 295.51*** \\

\rowcollight \multicolumn{8}{l}{\textbf{Syntax}}\\

Readability 
& 28.573 & 6.704 & 6.099
& \gradcell{-2.51}*
& \gradcell{-2.57}*
& \gradcell{-1.34}
& 284.88*** \\

Repeatability 
& 0.198 & 0.400 & 0.520
& \gradcell{27.64}***
& \gradcell{39.40}***
& \gradcell{22.59}***
& 976.60*** \\

Complexity 
& 8.274 & 3.911 & 3.749
& \gradcell{-2.87}**
& \gradcell{-2.98}**
& \gradcell{-1.92}
& 221.35*** \\

\rowcollight \multicolumn{8}{l}{\textbf{Style}}\\

Cat. Dynamic Idx. (CDI)
& -2.671 & 0.027 & -4.301
& \gradcell{2.90}**
& \gradcell{-1.69}
& \gradcell{-8.72}***
& 48.32*** \\

Empathy 
& 0.978 & 0.983 & 0.987
& \gradcell{5.80}***
& \gradcell{12.80}***
& \gradcell{7.18}***
& 234.29*** \\

Formality 
& 0.396 & 0.624 & 0.678
& \gradcell{15.31}***
& \gradcell{18.22}***
& \gradcell{5.39}***
& 324.56*** \\

Hope 
& 0.858 & 0.777 & 0.771
& \gradcell{-10.76}***
& \gradcell{-11.55}***
& \gradcell{-0.83}
& 202.64*** \\

\rowcollight \multicolumn{8}{l}{\textbf{Personalization}}\\

Ling. Accommodation
& 0.793 & 0.925 & 0.919
& \gradcell{14.42}***
& \gradcell{13.29}***
& \gradcell{-2.51}*
& 149.92*** \\

Semantic Similarity
& 0.289 & 0.445 & 0.394
& \gradcell{24.09}***
& \gradcell{14.42}***
& \gradcell{-8.88}***
& 328.34*** \\

Diversity
& 0.165 & 0.434 & 0.419
& \gradcell{41.74}***
& \gradcell{36.99}***
& \gradcell{-5.33}***
& 928.23*** \\

\bottomrule
\end{tabular}}
\caption{\textbf{\llum{}--IM}: Linguistic metric comparisons between human-written comments (OC), GPT baseline generations, and \llum{}-improved comments (IM). 
Pairwise comparisons are evaluated with paired $t$-tests, and overall differences are evaluated with Kruskal--Wallis $H$-tests (* $p$<0.05, ** $p$<0.01, *** $p$<0.001).
\vspace{-1em}
}
\label{tab:linguistic_improvement_comparison}
\end{table}

We next evaluate the effectiveness of the comment improvement model, which refines an initial human-written response given the original post. Detailed descriptions of all linguistic and psycholinguistic metrics, including their definitions, computation procedures, and interpretations, are provided in Appendix~\ref{sec:ling-analysis-explained}. As shown in Table~\ref{tab:linguistic_improvement_comparison}, \llum{}-improved comments show consistent gains over original comments and GPT baseline generations across multiple dimensions. 
In particular, improved comments show higher empathy and formality scores, alongside increased linguistic accommodation and diversity. 
At the same time, readability is substantially improved relative to original comments, suggesting that the model enhances clarity without sacrificing supportive tone. 
These results suggest that the improvement model effectively builds upon existing user-written content, refining it to better align with linguistic characteristics associated with effective peer support in online mental health communities.

\subsection{Human Evaluation for IM}

\begin{table}[t]
\centering
\scriptsize
\setlength{\tabcolsep}{3pt}
\renewcommand{\arraystretch}{0.9}
\begin{tabular}{lrrr}
\textbf{Model} 
& \textbf{Preference}
& \textbf{Content} 
& \textbf{Style} \\
\toprule

Original Response & 16.3\% of 424 & 2.83 & 2.38\\

GPT-5-nano
& 85.4\% of 213
& 3.10
& 2.71 \\

\llum{}--IM
& 82.0\% of 211
& 2.89
& 2.67\\

Either improvement model
& 83.7\% of 424
& 3.03
& 2.75\\

\bottomrule
\end{tabular}
\caption{\textbf{\llum{}--IM}: Participant ratings for original online community response (OC) and improvement model outputs across GPT-5-nano and \llum{}--IM, across dimensions of Content and Style match.}
\label{tab:revision_vote_results}
\vspace{-1em}
\end{table}

The human evaluations for IM model outputs consist of post--response pairs containing improved responses generated by either the GPT baseline or our improvement model. 
Participants are asked to choose which response they prefer (original or model-revised) and to evaluate whether the revised response preserves:
1) \textbf{Content}: The revised response feels consistent with the original response's content, and 2) \textbf{Style}: The revised response feels consistent with the original response's voice or style.
These are rated on the same 1 (\textit{Strongly disagree} to 5 (\textit{Strongly agree}) scale as before.

\autoref{tab:revision_vote_results} shows the proportion of participant preferences of responses. 
Overall, participants preferred the model-generated responses in 355 out of 424 comparisons (83.7\%), suggesting that both GPT and \llum{} were generally perceived as producing stronger supportive responses than the original Reddit comments. 
GPT responses were preferred in 182 out of 213 comparisons (85.4\%), while responses generated by the \llum{}--IM were preferred in 173 out of 211 comparisons (82.0\%), showing that \llum{} achieved a comparable preference rate, indicating that the model effectively improved response quality.

For content preservation, GPT-5-nano achieved a mean score of 3.103 while \llum{}--IM achieved 2.891, with a paired $t$-statistic of 1.036 but no statistical significance ($p$>0.05).
Similarly, for style preservation, GPT-5-nano achieved a score of 2.709 compared to 2.673 for \llum{}--IM, with a paired $t$-statistic of 0.014 and no statistical significance ($p$>0.05).
These findings suggest that the IM model preserves the original meaning, intent, and conversational style of user-written Reddit responses at a level comparable to GPT-5-nano while still improving supportive communication quality. 
\section{Discussion and Conclusion}
We studied how smaller, open-source language models can be adapted into effective writing assistants for online mental health support, by building \llum{} setup for both response generation and response improvement. 
Across linguistic analyses and human evaluations, \llum{} performed competitively with proprietary GPT.
Our results suggest that effective alignment for supportive communication does not depend only on model scale. 
Rather, community-specific data, preference supervision, and iterative human feedback can enable smaller models to provide emotionally appropriate and socially grounded writing assistance. 
Taken together, \llum{} demonstrates the potential of combining authentic online community interactions, community-derived preference signals, and human-centered alignment to build privacy-conscious AI systems that strengthen, rather than replace, peer support in online mental health communities.

\para{Community-aligned writing assistance.}
A key implication of this work is that AI systems for mental health peer support should be designed as \textit{writing assistants for human responders}, rather than as autonomous mental health agents. 
Online mental health communities often rely on moderators, volunteers, and everyday peers who want to help but may not have formal training in counseling, crisis communication, or emotionally sensitive writing. 
At the same time, these communities frequently struggle to train volunteers who can provide timely, high-quality, and safe responses at scale. 
In this context, tools such as \llum{} can help reduce the burden on peer responders by assisting them in drafting or revising comments to be clearer, warmer, more validating, and less likely to unintentionally cause harm. 
Importantly, this framing preserves the peer-driven nature of online support: the AI does not replace the community member, but helps them communicate more effectively.

\para{Implications for platform design.}
Our findings also point to concrete opportunities for designing future online support platforms. 
Rather than deploying AI as a direct responder to vulnerable users, platforms could embed writing assistance into the comment-composition interface. 
For example, when a user writes a response, the platform could offer optional revisions that improve empathy, clarity, validation, and safety while preserving the responder's original intent and voice. 
Such systems could also provide lightweight feedback before posting, such as flagging language that may sound dismissive, overly directive, generic, or potentially unsafe. 
This design direction shifts AI support from replacing human care to strengthening community care. 
It also creates a more accountable deployment model, where humans remain in control of what is posted and AI suggestions remain assistive rather than authoritative.

\para{The value and limits of community feedback.}
Our results further show that community-derived signals, such as upvotes and downvotes, can provide scalable supervision for training supportive response models. 
These signals are valuable because they emerge from real interactions and reflect how community members endorse, reject, or value different kinds of support. 
However, they should not be treated as perfect measures of supportive quality. 
Upvotes may reflect visibility, writing style, popularity, or what feels good to bystanders rather than what is most helpful to the person in distress. 
The improvement from additional DPO training with survey-derived human preferences suggests that community feedback is most useful when combined with direct human evaluation. 
Our work inspires that future models can treat community feedback as one layer of alignment, complemented by human-centered evaluation, expert review, and safety-aware moderation.

\para{Privacy-conscious and resource-sensitive deployment.}
This work also highlights the practical value of self-hosted open-source models for sensitive support contexts. 
Proprietary cloud-based APIs may require mental health-related disclosures to be transmitted to external providers, raising concerns about privacy, confidentiality, and data governance. 
In contrast, \llum{} can be deployed in-house within protected environments, giving platforms and community organizations greater control over data handling, auditing, and long-term deployment costs. 
This is especially important for resource-constrained communities that may not have access to large teams of trained volunteers, clinical experts, or expensive proprietary AI infrastructure. 
Smaller open-source models, when carefully aligned, may therefore offer a more feasible path for responsible AI-assisted support.
\section{Limitations and Future Directions}

Our work has several limitations which also suggest interesting future directions. First, although Reddit provides large-scale naturally occurring peer-support interactions, community feedback signals such as upvotes and downvotes are imperfect proxies for supportive quality and may also reflect factors such as popularity, writing style, or subreddit-specific cultural norms. Second, while our human evaluation study measures perceived empathy, supportiveness, and safety, it does not assess long-term psychological outcomes or the real-world impact of AI-assisted responses on users experiencing distress. In addition, our models are trained primarily on English-language Reddit communities, which may limit generalizability across cultures, languages, and support contexts. Future work could explore more diverse online communities, longitudinal evaluations of user wellbeing outcomes, and hybrid alignment approaches that combine community-derived feedback with expert clinical annotations. We also believe future systems could incorporate stronger safety-aware moderation mechanisms and personalized retrieval strategies to better adapt supportive responses to individual users and contexts.
\section{Ethical Considerations}

This work focuses on AI-assisted support for online peer-support communities and is not intended to replace professional mental health care, crisis intervention, or clinical counseling. Because the dataset is derived from publicly available Reddit discussions involving emotionally sensitive content, we take steps to minimize potential harm by removing personally identifying information and avoiding the release of raw user data. We additionally focus specifically on \textit{r/SuicideWatch} because it is a well-moderated mental health support community that has been extensively studied in prior work on online peer support and supportive communication. The community maintains relatively strong norms surrounding empathetic, non-judgmental, and safety-conscious responses, making it a more appropriate setting for studying supportive interaction patterns and community preference signals.

At the same time, we acknowledge that community feedback signals such as upvotes and downvotes are imperfect proxies for supportive quality and may reflect social biases, subreddit-specific norms, or collective preferences that do not always align with clinically appropriate support practices. More broadly, when applying community-driven preference alignment approaches, it is important to consider the moderation structure, behavioral norms, and values promoted within a given online community. In communities where harmful or toxic behaviors are normalized or highly rewarded, engagement-based preference signals could reinforce undesirable behaviors during model training. Future work should therefore carefully evaluate the social dynamics and moderation practices of online communities before adopting community-derived engagement signals for preference optimization.

Additionally, despite alignment efforts and human evaluation, generated responses may still produce inappropriate, unhelpful, or unsafe outputs in certain situations. For this reason, we position \llum{} as a supportive writing-assistance tool rather than an autonomous mental health advisor. Future deployment of such systems should incorporate additional safety mechanisms, human oversight, and clear communication regarding the limitations of AI-generated support.

Our study was approved by the Institutional Review Board (IRB) at our institution. 
Participants were recruited through Prolific and compensated according to platform standards and estimated study completion time. 
Prior to participation, all participants were presented with informed consent materials describing the purpose of the study, the emotionally sensitive nature of the content, and how their responses would be used for research purposes(\autoref{prolific-instruction}). 
Because the study involves mental health-related Reddit discussions, participants were also informed that they could discontinue participation at any time without penalty. 
All collected evaluation data were anonymized and used solely for research purposes.

\section{AI Involvement Disclosure}
We used AI-assisted writing tools (e.g., ChatGPT and Grammarly) to refine and edit the writing of the manuscript. All analyses, scientific content, and experiments were written solely by the authors.


\section*{Acknowledgments}
This work was supported in part by the Jump ARCHES endowment through the Health Care Engineering Systems Center at the University of Illinois and the OSF Foundation. 
We thank Yunhao Yuan for contributing to the initial discussions and brainstorming for this work.

\bibliography{0paperACL}

\appendix
\clearpage
\appendix
\section{Appendix}
\setcounter{table}{0}
\setcounter{figure}{0}
\renewcommand{\thetable}{A\arabic{table}}
\renewcommand{\thefigure}{A\arabic{figure}}



\subsection{Linguistic Analysis Metrics}
\label{sec:ling-analysis-explained}
\subsubsection{Verbosity}
We measure response verbosity using \textbf{Words Per Response (WPR)} and \textbf{Words Per Sentence (WPS)}~\cite{saha2025ai}. WPR is computed as the average number of words contained in a complete response, while WPS measures the average sentence length within a response. These metrics characterize how much information a model provides and the level of elaboration in generated text. Higher WPR values indicate more detailed and expansive responses, although excessively large values may also reflect unnecessary verbosity. Higher WPS values suggest longer and potentially more complex sentence structures, whereas lower values indicate shorter and simpler expressions.

\subsubsection{Syntax}

\textbf{Readability} measures how easily a text can be understood by readers. Prior work has identified readability as an important characteristic in health communication and online support settings, where communication quality depends not only on what information is conveyed but also on how easily users can interpret and process it~\cite{mcinnes2011readability,ernala2017linguistic, saha2020causal}. We measure readability using the Coleman--Liau Index (CLI)~\cite{coleman1975computer}, a readability metric based on character and sentence structure: 
CLI = 0.0588L - 0.296S - 15.8
 where \(L\) denotes the average number of letters per 100 words and \(S\) denotes the average number of sentences per 100 words. Unlike readability measures based on syllable counts, CLI estimates text difficulty using word and sentence composition. Higher CLI values generally indicate more sophisticated writing and greater linguistic complexity; however, they also suggest that a higher level of education may be required to understand the text. 

 \textbf{Repeatability} and \textbf{complexity} are syntactic measures that characterize the richness and depth of language expression and have been associated with cognitive processes involved in communication, such as planning, execution, and information organization~\cite{ernala2017linguistic}. Repeatability captures how frequently words are reused within a response, while complexity reflects the sophistication and depth of language construction.

Following prior work~\cite{ernala2017linguistic,saha2020causal,yuan2023mental}, we measure repeatability as the normalized occurrence of non-unique words within a response. Higher repeatability values indicate greater lexical redundancy and repeated information, which may suggest reduced conciseness and lower informational crispness.

We measure complexity as the average word length per sentence~\cite{saha2025ai}. Higher complexity values generally indicate more sophisticated and nuanced language that may convey ideas with greater precision and depth. However, excessively high complexity can also increase cognitive burden and make text more difficult to understand.

\subsubsection{Style}

The \textbf{Categorical Dynamic Index (CDI)} measures linguistic style along a spectrum ranging from \textit{categorical} to \textit{dynamic} language~\cite{pennebaker2014small}. Categorical language reflects a more analytical and structured style that focuses on organized concepts and logical reasoning, whereas dynamic language is more narrative and socially oriented, often emphasizing personal experiences and storytelling. Prior work has shown that linguistic style is an important characteristic in social support and therapeutic communication contexts~\cite{cutrona1986social,saha2020causal}.

Following Pennebaker et al., CDI is computed as:

\[
\begin{aligned}
CDI =\ &30 + article + preposition \\
&- personal\ pronoun - impersonal\ pronoun \\
&- auxiliary\ verb - conjunction \\
&- adverb - negation
\end{aligned}
\]

where each term represents the percentage occurrence of the corresponding linguistic category within a text. We obtain these occurrences using the LIWC lexicon~\cite{tausczik2010psychological}. CDI is a bipolar measure in which higher values indicate a more categorical and analytical writing style, while lower values indicate a more dynamic, personal, and narrative style of communication.

\textbf{Empathy} is a central component of supportive communication and has been identified as an important mechanism in mental health support and social interactions~\cite{sharma2020computational}. Empathy refers to the ability to understand and recognize another individual's emotional state, experiences, and perspective, often through expressions of emotional understanding and validation~\cite{herlin2016dimensions}. In online support contexts, empathetic responses have been associated with more positive user perceptions and improved interaction quality~\cite{sharma2020computational,morris2018towards}.

To measure empathy, we use a RoBERTa-based empathy detection model fine-tuned on datasets of empathetic reactions to news stories~\cite{tafreshi2021wassa,buechel2018modeling}. Given a response, the model predicts an empathy score representing the degree to which the text expresses emotional understanding and supportive intent. Higher scores indicate stronger expressions of empathy and greater use of linguistic cues associated with emotional acknowledgment and perspective-taking.

\textbf{Formality} is an important sociolinguistic characteristic that reflects how language varies across audiences, contexts, and communication settings~\cite{heylighen1999formality,levin1991frequencies}. Formality refers to the degree to which text follows established linguistic conventions, including structured syntax, appropriate vocabulary, and adherence to grammatical norms. Formal language is commonly associated with professional and academic communication, whereas informal language often contains conversational expressions, colloquialisms, and more casual writing styles.

To measure formality, we use a RoBERTa-based formality classifier proposed in prior work~\cite{babakov2023don}. Given a text, the model predicts a formality score, where higher scores indicate greater use of formal language characteristics, while lower scores indicate a more conversational and informal writing style.

\textbf{Hopefulness} reflects the extent to which language conveys optimism, encouragement, and positive expectations about future outcomes. In mental health and supportive communication settings, expressions of hope can play an important role in fostering emotional support and promoting positive outlooks during distressing situations~\cite{sharma2020computational}. Supportive responses that communicate hope may encourage resilience and provide reassurance while avoiding excessively negative or fatalistic language.

To measure hopefulness, we use a zero-shot classification approach using the \texttt{facebook/bart-large-mnli} model proposed in prior work \cite{kyritsis2023zero}. BART-large-MNLI  is  a  variant  of  the  BART  model  fine-tuned specifically for the MNLI task. We frame hope detection as a natural language inference task with two candidate labels: \textit{hopeful} and \textit{not hopeful}. Given a response, the model estimates the probability that the text semantically entails the label \textit{hopeful}. Higher values indicate greater confidence that the text expresses hopeful language, while lower values indicate weaker expressions of optimism or encouragement.

\subsubsection{Personalization}

\textbf{Linguistic style accommodation} measures the extent to which a response adapts to the linguistic style of the original query. Prior work has shown that greater accommodation between individuals is associated with more effective communication and improved outcomes in online support interactions~\cite{saha2020causal,sharma2018mental}. Unlike content-based measures, accommodation focuses on stylistic characteristics of language rather than semantic content, capturing how individuals adapt their communication patterns through grammar and function word usage.

Following prior work~\cite{saha2025ai}, we measure linguistic accommodation by computing the cosine similarity between linguistic style vectors derived from the original query and its corresponding response. The vectors are constructed using normalized occurrences of non-content linguistic categories obtained through the LIWC lexicon~\cite{pennebaker2003psychological, saha2020causal}, including articles, prepositions, pronouns, auxiliary verbs, conjunctions, adverbs, and negations:

\[
Accommodation(q,r)
=
\frac{L(q)\cdot L(r)}
{||L(q)|| \ ||L(r)||}
\]

where \(L(q)\) and \(L(r)\) represent the linguistic style vectors of the query and response, respectively.

Higher accommodation values indicate stronger stylistic adaptation to the user's language patterns, suggesting greater alignment with the way users communicate. Lower values indicate weaker adaptation and a more generic response style.

\textbf{Semantic similarity} measures the degree to which a response remains semantically and topically aligned with the original query. In supportive communication settings, maintaining semantic relevance is important to ensure that responses address the user's concerns and remain coherent with the context of the discussion. Higher semantic similarity suggests that responses are more closely related to the themes and meaning expressed in the original post.

Following prior work~\cite{saha2025ai}, we measure semantic similarity by computing the cosine similarity between vector representations of the query and response. We obtain sentence-level embeddings using pretrained BERT-based sentence transformers~\cite{reimers2019sentence}. Semantic similarity is computed as:

\[
Sim(q,r)
=
\frac{E(q)\cdot E(r)}
{||E(q)|| \ ||E(r)||}
\]

where \(E(q)\) and \(E(r)\) denote the embedding representations of the query and response, respectively.

Higher similarity scores indicate stronger topical and semantic alignment between a query and its response, suggesting that the response more closely addresses the user's original concerns. Lower scores indicate weaker semantic coherence and potentially less relevant responses.

\textbf{Diversity} measures the degree of variation and uniqueness in generated responses. Prior work has shown that diverse and creative responses can improve the effectiveness of psychotherapy and supportive interactions by reducing repetitive language and enabling a broader range of supportive strategies~\cite{althoff2016large,norcross2018psychotherapy,saha2020causal}. In conversational systems, low diversity may indicate templated or repetitive responses, whereas greater diversity can reflect richer and more flexible language generation.

Following prior work~\cite{mikolov2013distributed, saha2025ai}, we measure diversity using word embedding representations in a 300-dimensional semantic space. For each dataset, we first compute a centroid vector representing the average embedding across all responses. We then calculate the cosine distance between each response embedding and its corresponding dataset centroid:

\[
Diversity(r)
=
1
-
\frac{E(r)\cdot C}
{||E(r)|| \ ||C||}
\]

where \(E(r)\) denotes the embedding representation of response \(r\), and \(C\) represents the centroid embedding of the dataset. Larger distance values indicate that a response deviates more from the average response representation and is therefore considered more diverse.

Higher diversity scores indicate greater linguistic variation and less repetitive or templated language, while lower values indicate responses that are more similar to common patterns within the dataset.

\subsection{Model Training Prompts}

\subsubsection{GM \sft{} Prompt (Instruction-tuning)}
\label{sec:gm-sft-prompt}
 
\begin{promptbox}
<s>[INST]  
``You are a Reddit comment writing assistant. Your goal is to write a supportive, empathetic, and helpful comment for the given Reddit post. Respond with warmth, validation, and concise support without repeating or over-empathizing. 

The post: {post\_text}.''
[/INST] 
{response}
</s>
\end{promptbox}

\subsubsection{GM \dpo{} Prompt}
\label{sec:gm-dpo-1-prompt}
 
\begin{promptbox}
You are a Reddit comment writing assistant. Your goal is to write a supportive, empathetic, and helpful comment for the given Reddit post. Respond with warmth, validation, and concise support without repeating or over-empathizing.

The post: {post_text}.
\end{promptbox}

\subsubsection{GM \dpoo{} Prompt}
\label{sec:gm-dpo-2-prompt}
 
\begin{promptbox}
You are a Reddit comment writing assistant. Your goal is to write a supportive, empathetic, and helpful comment for the given Reddit post. Respond with warmth, validation, and concise support without repeating or over-empathizing.

The post: {post_text}.
\end{promptbox}

\subsubsection{IM \sft{} Training Prompt using RAG}
\label{sec:im-rag-prompt}
 
\begin{promptbox}
<s>[INST]
<role>
You are a Reddit commenting assistant. Your role is to read the original <post> and the user's <initial_comment>, then improve that <initial_comment> to make it sound more supportive, empathetic, and helpful toward the original poster.
</role>

<commenting_guidelines>
1. **Tone and Style**
- Sound like a genuine Reddit user---warm, human, and conversational.
- Avoid overly formal, clinical, or ``therapist-like'' language.
- Do not use generic or repetitive empathy (e.g., ``I'm so sorry you're going through this'' repeatedly).
- Avoid using emojis. 
- Do not suggest any type of hotline or crisis resource in your response (e.g., ``If you're in immediate danger or think you might harm yourself again, call emergency services or go to the nearest ER. If you're in the US: call or text 988, or chat at 988lifeline.org.'').

2. **Empathy and Support**
- Show understanding and validation of the original poster's feelings.
- Offer encouragement, gentle advice, or shared perspective where appropriate.
- Be kind, authentic, and emotionally intelligent---not robotic or exaggerated.

3. **Content Quality**
- Improve clarity, flow, and emotional resonance.
- Keep it **concise**---avoid long, repetitive paragraphs.
- Ensure the improved comment provides **constructive emotional or practical support**.
- Maintain or enhance any useful information from the original comment.
</commenting_guidelines>

<examples>
Post: [Retrieved similar Reddit post]
Supportive Comment: [Retrieved supportive response]

---

Post: [Retrieved similar Reddit post]
Supportive Comment: [Retrieved supportive response]
</examples>

<post>
{post}
</post>


<initial_comment>
{init_comment}
</initial_comment>
[/INST]{target_comment}
</s>
\end{promptbox}
\subsubsection{Retriever Training Objective}
\label{sec:retriever-prompt}
 
\begin{promptbox}
Query:
{post_title} {post_body}

Positive Example:
{generated_target_comment}

Negative Example:
{initial_comment_body}
\end{promptbox}

\subsubsection{IM: GPT Prompt for Generating Revised Comments}
\label{sec:gpt-data-prompt}
 
\begin{promptbox}
<role>
You are a Reddit commenting assistant. Your role is to read the original <post> and the user's <initial_comment>, then improve that <initial_comment> to make it sound more supportive, empathetic, and helpful toward the original poster.
</role>

<commenting_guidelines>
1. **Tone and Style**
- Sound like a genuine Reddit user---warm, human, and conversational.
- Avoid overly formal, clinical, or ``therapist-like'' language.
- Do not use generic or repetitive empathy (e.g., ``I'm so sorry you're going through this'' repeatedly).
- Avoid using emojis. 
- Do not suggest any type of hotline or crisis resource in your response (e.g., ``If you're in immediate danger or think you might harm yourself again, call emergency services or go to the nearest ER. If you're in the US: call or text 988, or chat at 988lifeline.org.'').

2. **Empathy and Support**
- Show understanding and validation of the original poster's feelings.
- Offer encouragement, gentle advice, or shared perspective where appropriate.
- Be kind, authentic, and emotionally intelligent---not robotic or exaggerated.

3. **Content Quality**
- Improve clarity, flow, and emotional resonance.
- Keep it **concise**---avoid long, repetitive paragraphs.
- Ensure the improved comment provides **constructive emotional or practical support**.
- Maintain or enhance any useful information from the original comment.
</commenting_guidelines>

<response_format>
- Output only the **improved comment**, nothing else (no explanations, no markdown, no quotes).
</response_format>

<post>
{post_content}
</post>

<initial_comment>
{initial_row.comment_body}
</initial_comment>
\end{promptbox}

\subsection{Data Validation for IM}

\begin{table}[H]
\centering
\small
\resizebox{\columnwidth}{!}{
\begin{tabular}{lccc}
\textbf{Metric} & \textbf{Initial Comment} & \textbf{Generated Comment (\llum{})} & \textbf{$t$-test Significant} \\
\toprule
cdi & -3.49 & 2.89 & \textbf{Yes} \\
complexity & 3.97 & 4.09 & \textbf{Yes} \\
diversity\_sent & 0.23 & 0.44 & \textbf{Yes} \\
empathy & 0.97 & 0.99 & \textbf{Yes} \\
formality & 0.57 & 0.59 & No \\
hope & 0.85 & 0.78 & \textbf{Yes} \\
ling\_style & 0.88 & 0.94 & \textbf{Yes} \\
readability & 4.84 & 7.95 & \textbf{Yes} \\
repeatability & 0.26 & 0.43 & \textbf{Yes} \\
wpr & 53.61 & 124.27 & \textbf{Yes} \\
wps & 13.96 & 15.72 & \textbf{Yes} \\
\bottomrule
\end{tabular}}
\caption{Paired linguistic analysis comparing initial human-written comments and \llum{}-generated improved comments. Statistical significance is determined using independent $t$-tests ($p < 0.05$).}
\label{tab:paired_linguistic_significance}
\end{table}

\begin{table}[H]
\centering
\small
\resizebox{\columnwidth}{!}{
\begin{tabular}{llc}
\textbf{Category} & \textbf{Metric} & \textbf{Mean Score} \\
\toprule
\multirow{1}{*}{Embedding Similarity} 
& Cosine similarity (Initial vs.\ Improved) & 0.244 \\
\midrule
\multirow{3}{*}{Cross-Encoder STS} 
& STS (Initial vs.\ Post) & -1.914 \\
& STS (Improved vs.\ Post) & -1.203 \\
& STS (Improved vs.\ Initial) & -0.613 \\
\midrule
\multirow{4}{*}{NLI Consistency} 
& Entailment (Initial $\rightarrow$ Improved) & 0.154 \\
& Contradiction (Initial $\rightarrow$ Improved) & 0.032 \\
& Entailment (Improved $\rightarrow$ Initial) & 0.226 \\
& Contradiction (Improved $\rightarrow$ Initial) & 0.160 \\
\midrule
\multirow{2}{*}{BERTScore} 
& BERTScore-F1 (Improved vs.\ Initial) & 0.835 \\
& BERTScore-F1 (Improved vs.\ Post) & 0.794 \\
\bottomrule
\end{tabular}}
\caption{Semantic similarity and consistency analysis between initial comments and their \llum{}-improved versions. Higher similarity and entailment scores indicate stronger content preservation, while lower contradiction scores indicate reduced semantic divergence.}
\label{tab:similarity_analysis}
\end{table}

\subsection{Prolific Instructions}
\label{prolific-instruction}
 
\begin{promptbox}
Study name: Evaluating Supportive Responses in Online Mental Health Communities

Study Description: Participants will complete an online survey in which they review a set of short post-response examples from online mental health peer-support contexts. Each example consists of a post and one or more responses. For each example, participants will answer structured questions evaluating the responses (e.g., perceived supportiveness, empathy, and helpfulness). Approximately 10 examples will be evaluated. The survey is self-paced and requires no writing or sharing of personal experiences.
\end{promptbox}

\end{document}